\documentclass[acmsmall]{acmart}
\AtBeginDocument{%
  \providecommand\BibTeX{{%
    \normalfont B\kern-0.5em{\scshape i\kern-0.25em b}\kern-0.8em\TeX}}}

\setcopyright{acmlicensed}
\acmJournal{PACMHCI}
\acmYear{2023} \acmVolume{7} \acmNumber{CSCW2} \acmArticle{TBD} \acmMonth{11} 
\acmPrice{15.00}\acmDOI{XXXXX.XXXXX}

\acmConference[CSCW '23]{Minneapolis '23: ACM Conference On Computer-Supported Cooperative Work And Social Computing}{October 13--18, 2023}{Minneapolis, MN USA}
\acmPrice{15.00}
\acmISBN{978-1-4503-XXXX-X/18/06}

\acmSubmissionID{7000}

\usepackage{fontawesome,dirtytalk}
\usepackage{xcolor}
\usepackage{graphicx,multirow,array}
\usepackage{color,soul}
\usepackage{tikz}
\usepackage{libertine}
\usepackage{rotating,tabularx}

\begin{document}

\title[User Perspectives on Branching in Computer-Aided Design]{User Perspectives on Branching in Computer-Aided Design}


\author{Kathy Cheng}
\affiliation{%
  \institution{University of Toronto}
  \streetaddress{27 King's College Circle}
  \city{Toronto}
  \country{Canada}}
\email{kathy.cheng@mail.utoronto.ca}

\author{Phil Cuvin}
\affiliation{%
  \institution{University of Toronto}
  \city{Toronto}
  \country{Canada}}

\author{Alison Olechowski}
\affiliation{%
  \institution{University of Toronto}
  \city{Toronto}
  \country{Canada}}

\author{Shurui Zhou}
\affiliation{%
  \institution{University of Toronto}
  \city{Toronto}
  \country{Canada}}

\renewcommand{\shortauthors}{Cheng et al.}

\begin{abstract}

Branching is a feature of distributed version control systems that facilitates the ``divide and conquer'' strategy present in complex and collaborative work domains. Branching has revolutionized modern software development and has the potential to similarly transform hardware product development via CAD (computer-aided design). Yet, contrasting with its status in software, branching as a feature of commercial CAD systems is in its infancy, and little research exists to investigate its use in the digital design and development of physical products. To address this knowledge gap, in this paper, we mine and analyze 719 
user-generated posts from online CAD forums to qualitatively study designers’ intentions for and preliminary use of branching in CAD. Our work contributes a taxonomy of CAD branching use cases, an identification of deficiencies of existing branching capabilities in CAD, and a discussion of the untapped potential of CAD branching to support a new paradigm of collaborative mechanical design. The insights gained from this study may help CAD tool developers address design shortcomings in CAD branching tools and assist CAD practitioners by raising their awareness of CAD branching to improve design efficiency and collaborative workflows in hardware development teams.


\end{abstract}



\maketitle

\section{Introduction} 
Hardware design and development -- the creation of physical products -- is an indispensable part of human history and essential to technological innovation. Contemporary hardware development is practically impossible without the use of CAD (computer-aided design) software to facilitate all aspects and stages of the development process, from initial conceptual design to manufacturing of the final product~\cite{Stark2022}. 

The field of hardware development is long-standing, as are the tools and systems involved in the hardware design process. Traditional CAD ecosystems such as PLM (Product Lifecycle Management), PDM (Product Data Management), and centralized version control that first appeared in the 1980s remain the standard collaboration systems in place at most hardware organizations today~\cite{Chou1986,Maher2017,Warfield2021}. As design practices are becoming increasingly globalized, distributed, and demanding~\cite{Bykzkan2011}, these traditional ecosystems are unable to keep up with the changing nature of hardware design and development work~\cite{Cheng2023}. Recent advances in new generation collaboration and version control tools -- such as cloud-based CAD, fully-synchronous CAD, and distributed version control (i.e., branching and merging) -- have the potential to support today's complex collaborative workflows. Similar tools have become second nature to the related field of software development, fundamentally changing the way in which software developers collaborate.


Branching and merging is a feature from distributed version control systems (e.g., Git\footnote{https://git-scm.com/}~\cite{Otte2009}, Mercurial\footnote{https://www.mercurial-scm.org/}) popularized by cloud-based code hosting platforms (e.g., GitHub, GitLab, Bitbucket), that is essential to modern distributed version control systems (DVCS) \cite{Brindescu2014,Phillips2011,Muslu2014}. Branching makes it possible for developers to split a codeline away from the production code version to make contributions without affecting the existing version or disturbing collaborators unnecessarily~\cite{Zou2019,Zhou2018}, while merging allows changes made in ``source branches'' to be integrated into a single ``target branch'' or ``main branch''~\cite{Phillips2011}. Though branching and merging are related operations, in this work, we focus on branching practices due to the limited abilities of current CAD merging functionalities
(as we discuss later in this paper) and the fact that merging is a consequence of branching; thus, it is in our best interest to first study users' intentions for branching. Related work in software literature likewise prioritizes thorough understanding of branching practices before further investigation of merging~\cite{Shihab2012, Phillips2011}.
\color{black}
Branching and merging functionality has revolutionized modern software development; collaboration has not only become more efficient and effective, but a new paradigm of open-source and crowdsourced software development work has been facilitated with branching tools~\cite{Zou2019,O'Sullivan2009}.

Although hardware and software developers create different products (physical vs. digital), both fields are complex, large-scale, and inherently collaborative work domains which require the collective and joint contribution of many individuals~\cite{Schmidt2013}. Both fields are challenged by collaborative distance, overhead of synchronization, and lack of group awareness~\cite{dourish1992awareness,olson2000distance,Cheng2023}, which necessitates version control systems to manage file documentation, track version changes, and enable access to shared files~\cite{Bricogne2012,pete2015handling}. While there are parallels between software and hardware development, artifacts in CAD are 3D models that represent geometric and topological data (e.g., \emph{STL}, \emph{STEP} files), which are fundamentally different from artifacts in software, which consist primarily of text-based code and documentation~\cite{Sun2010,Stirling2022}. Therefore, while lessons from software development are informative to CAD researchers and users, their fundamental differences motivate a dedicated study for hardware branching. 
\color{black}

To date, little research has investigated CAD branching for the design and development of hardware products. Existing work has focused on proposing computational or data management methods for accomplishing branch and merge operations (e.g., hash algorithms \cite{Fresemann2021}, ``diff and merge'' using persistent identifiers (POID) \cite{Richter2006}, and processing-oriented modeling \cite{Koch2011}), but few studies have focused on CAD and CAD-related systems from the user perspective \cite{Barrios2022}. Furthermore, support for branching and merging in commercial CAD platforms is a relatively recent development. Onshape\footnote{https://www.onshape.com/en/} was the first to bring this capability to CAD artifacts in 2015, followed by Fusion360\footnote{https://forums.autodesk.com/} in 2017, and SolidWorks PDM\footnote{https://www.solidworks.com/product/solidworks-pdm} in 2018~\cite{onshape2015,medeiros2017}. \color{black} However, the implementation of these tools has not always been successful, for example, commercial releases from Fusion360 were discontinued soon after launch~\cite{song2017}. As branching technology is still relatively new to the CAD community, this exploratory study is needed to uncover how branching is currently used and adopted among CAD designers, what the pain points are of current tools, and how to improve branching functionalities to fulfill the unmet design needs of hardware designers. 


Our study aims to close this knowledge gap through an empirical investigation of designers’ current intentions and expectations for the use of branching in CAD. The following two questions guided our research:

\begin{itemize}
  \item[] \textbf{RQ1:} How do CAD designers use branching functionality in the hardware design process? 
  \item[] \textbf{RQ2:} What are the design shortcomings and gaps of existing CAD branching tools?
\end{itemize}

To answer these research questions, we mine and qualitatively analyze 719 user-generated posts from online CAD forums, inclusive of CAD-platform-specific forums (i.e., Fusion360\footnote{https://forums.autodesk.com/}, Onshape\footnote{https://www.onshape.com/en/}, and SolidWorks PDM\footnote{https://www.solidworks.com/product/solidworks-pdm}), and platform-agnostic forums (i.e., CAD Forum\footnote{https://cadforum.net/} and Eng-Tips\footnote{https://www.eng-tips.com/}). \color{black} We chose to mine posts from online CAD forums for our exploratory study because searching in online forums allows us to reach more CAD users and obtain more discussions of branching than we would with other qualitative research methods, such as interviews. Furthermore, online communities which communicate in user forums provide a rich and comprehensive dataset that can help researchers understand real user requirements and behaviours; other CSCW (Computer-Supported Cooperative Work) studies have similarly mined online forums to understand the privacy concerns of software developers~\cite{Li2021}, misinformation in social media platforms~\cite{Jiang2018}, and codes of conduct in collaborative software projects~\cite{Li2021code}.
\color{black}


We therefore contribute the following insights to the growing CAD community within CSCW:
\begin{enumerate}
    \item An empirical investigation of 719 user-generated online  posts from five forum sites to identify shortcomings of existing branching capabilities in commercial CAD platforms, providing tool builders with actionable improvements for CAD branching tools.
    \item A taxonomy of CAD branching use cases, comprised of three high-level groupings, \emph{Product Line Management}, \emph{Risk Isolation}, and \emph{Designer Support}.
    \item Evidence that CAD designers use branches for several use cases within each grouping, most frequently for production version maintenance, creating new product variants, and presenting the design to project stakeholders. Our findings further revealed that not all identified use cases are intentional functions of branching tools; in other words, branching in CAD is often used as a workaround to perform other desired design tasks -- potentially due to lack of technical support in these areas. 
    \item A discussion of the potential of branching to not only support a new paradigm of collaborative mechanical design, but improved coordination, cooperation, and communication for CSCW domains as a whole. 
\end{enumerate}





\section{Background \& Related Work}\label{sec_background}


\subsection{Branching and Merging for Software Development}\label{sec_branchingsoftware}

In software development, branching is a mechanism supported by distributed version control systems (e.g., Git, Mercurial) that allows for a code file to be cloned and edited independently of the main codeline, and to subsequently be merged to the original file to reconcile the changes with the main codeline~\cite{Phillips2011}. Branching and merging is currently the most widely adopted method for software version control~\cite{Phillips2011, Zou2019, Bird2012}. 

Branching functionality has been present in software version control tools since the 1980s~\cite{Tichy1985}, but modern branch-based distributed version control only entered wider use with the introduction of Git (an open-source version control tool which provided lightweight and robust branching and merging workflows) in 2005, and with the introduction of GitHub in 2008, which leveraged branching and merging functionality derived from Git to enable large-scale social collaboration on software projects and to provide a platform for open-source development~\cite{Phillips2011, Zou2019, O'Sullivan2009}. 

Branching enables parallel and simultaneous collaboration on the same code file and offers developers the ability to isolate the risk associated with changing code (and potentially breaking the program) from the main branch (also known as the ``mainline'' or the ``root'' branch)~\cite{Microsoft2022, Zou2019}. Therefore, one might branch for bug fixes, new feature development, experimental code creation, and maintenance of build scripts~\cite{Phillips2011, O'Sullivan2009}. Branching is also used to support organizational and non-technical workflow requirements, such as file sharing or dividing work between programming teams~\cite{Phillips2011}.

\color{black}

\subsubsection{Purposes of Branching in Software Development}
Software version control branches often fulfill more than a single purpose, as a branch can accomplish multiple development objectives at the same time (e.g., a branch can simultaneously isolate a bug fix from the root branch and assign the bug fix task to a separate team)~\cite{Phillips2011, Zou2019}. As such, a use case, as we will present, represents a purpose or intention of branching that does not overlap with any other. Given this definition of a branching use case, we synthesized prior literature that explored how software developers use branching in practice, in order to provide an overview of known software branching use cases. 

%

Previous studies on software branching have demonstrated that developers primarily use branches for version control purposes (e.g., tracing design history, tracking version changes, managing product evolution) and this is consistent with how branches were intended to be used, as described in foundational handbooks and guides~\cite{Wingerd1998,Appleton1998,Walrad2002}. The most popular use cases discussed in literature are branching to separate product releases for different devices or versions of operating systems (i.e., \emph{Parallel Version} branching), and branching to build new features that are intended to be integrated into the main branch and put into production (i.e., \emph{New Feature} branching)~\cite{Phillips2011,Zou2019}. Creating a branch per feature is a development strategy that presents the opportunity for \emph{Modularization} branching, where branches separate code into modules that can be combined to result in different product configurations~\cite{Murphy2014}. Prior work has also shown that developers use branches to build faster, better, and more optimized versions of the mainline code, without altering the function of the code, known as \emph{Refactorization} branching~\cite{Zou2019,O'Sullivan2009}. An important theme that emerges from these studies is that the aforementioned use cases serve the development, iteration, and evolution of the mainline branch, or production code version. 

In practice however, developers not only use branching to serve the mainline branch, but also to isolate the risk of unverified and untested changes from the mainline; such branches are described as ``small-scope, short-lived, and used for single, targeted purposes''~\cite{Phillips2011} and ``more temporary and for minor changes''~\cite{Walrad2002}. For example, bug fixing commonly requires developers to create these ``ad-hoc'' branches, where an error in the code must be repaired and merged back~\cite{Phillips2011,Zou2019,Murphy2014,Premraj2011}. Developers may branch to experiment with building new code that is not yet confirmed as a feature (i.e., \textit{Experimental/Prototype} branching)~\cite{Phillips2011}, or branch to create an isolated environment for testing code changes prior to integration with the mainline (i.e., \emph{Testing} branching)~\cite{Phillips2011,Zou2019,Murphy2014}. Branches are further used to maintain and store process-specific code which are not directly related to the main codebase, but are necessary for building, testing, or deploying software (e.g., build scripts, macros, regression test suites) outside of the main development branch (i.e., \emph{Tool/Utility} branching)~\cite{Phillips2011}.


Software literature has also identified that branching can be used to serve developers' organizational or non-technical needs~\cite{Premraj2011,Phillips2011}. A frequently discussed purpose of branching is to support collaboration by dividing coding work between different individuals or teams (i.e., \emph{Work Division} branching)~\cite{Premraj2011,Phillips2011,Appleton1998}, or allowing shared files to be easily accessed by multiple developers (i.e., \emph{File Sharing} branching)~\cite{Appleton1998,Premraj2011,Wingerd1998,Bird2011}. It has been shown that the parallel workflow made possible by branching can improve working efficiency in collaborative software development projects~\cite{Shihab2012,Premraj2011,Murphy2014}.

Overall, prior work has identified numerous branching use cases for software development, summarized in Table~\ref{softwarebranch}. These studies provide an important foundation for our knowledge of software branching and have high potential to inform our understanding of how branches may be used in other contexts, like the development of hardware products. Nonetheless, compared to software branching, relatively little is known about CAD branching purposes in the hardware development world, therefore motivating the present study. 


\begin{table}[ht]
\footnotesize
  \caption{Summary of software branching use cases identified from literature.}
  \begin{center}
  \label{softwarebranch}
  \begin{tabular}{ll p{2.1in} l}
  \hline
    Grouping&Use Case&Description&Reference\\
    \hline
    Product Line Management &Parallel Version&Separate different product configurations or contexts (e.g., x86 vs. x64).&\cite{Phillips2011,Zou2019}\\
    & Modularization &Partition code into components that can be combined to create various configurations. &\cite{Murphy2014}   \\
    & Refactorization &Perform code refactoring or structure optimization.    &\cite{Phillips2011,Zou2019,Premraj2011}  \\
    &New Feature &Implement a new feature(s).    &\cite{Phillips2011,Zou2019,Murphy2014}    \\
    \hline
    Risk Isolation & Tool/Utility & Maintain process-specific code (e.g., build scripts, macros, and regression test suites).  &\cite{Phillips2011}      \\
    &Testing&Test code changes prior to integration to the main branch.   &\cite{Phillips2011,Zou2019}   \\
    &Experimental/Prototype&Isolate disruptive or unstable code that may not go into production. &    \cite{Premraj2011,O'Sullivan2009,Phillips2011}    \\
    &Bug Fix & Resolve bugs, errors, or problems with the code.   &\cite{Phillips2011,Zou2019,Murphy2014}    \\
    \hline
    Developer Support &Work Division&Isolate individual work before it is shared with the product team.&\cite{Premraj2011,Phillips2011,Appleton1998}\\
    &File Sharing &Coordinate the distribution of files among collaborators. &\cite{Premraj2011,Phillips2011,Bird2011}  \\
  \hline
\end{tabular}
\end{center}
\end{table}



\color{black}
\subsection{Branching and Merging for CAD}

Branching and merging as a method for CAD version control is not a new concept in theory; literature as early as the 1980s has discussed distributed version control systems for CAD~\cite{Chou1986, Katz1990,Liu1990}. 
However, the practical branching of CAD artifacts in commercial platforms is a relatively recent development -- to our knowledge, support for branching and merging of parametric CAD files has emerged in only three CAD platforms: PTC's Onshape in 2015~\cite{onshape2015}, Autodesk's Fusion360 in 2017 (soon after removed as functionality, also in 2017)~\cite{song2017}, and Dassault Systèmes's SolidWorks PDM in 2018~\cite{medeiros2017}. 
While branches for CAD artifacts (e.g., 3D models, CAD drawings) have been supported in commercial CAD tools since 2015 (by Onshape), the investigation of practical usage of branching and merging in CAD literature remains scarce.

Richter and Beucke explored version control for distributed civil engineering design and proposed ``diff and merge'' solutions inspired by the software development field to compare and merge versions of CAD drawings~\cite{Richter2006}. They identified that the ``parallel and consistent collaboration'' that is enabled by branching can be a superior workflow compared to traditional centralized document management systems. Le performed a SWOT (strength, weakness, opportunity, threat) analysis to compare the CAD platforms, Onshape and SolidWorks and concluded that one of the main reasons why users preferred Onshape is the flexibility of its branch and merge tool that allows users to model concurrently as well as experiment with multiple design variations~\cite{Le2018}. Stirling et al. explored the possibility of adapting DevOps workflows -- widely utilized in software development -- to facilitate open collaborative development of hardware products, creating ``HardOps''~\cite{Stirling2021,Stirling2022}. In open-source hardware projects, it is essential that multiple designers can simultaneously contribute to a design and for all contributors to have unfettered access to all product data, and this cannot be achieved with the centralized version control systems in place at most design organizations~\cite{Stirling2022}. With distributed (i.e., branch-based) version control, contributors to open hardware projects can work independently, test and refine their changes, and ensure that their changes are carefully reviewed before being integrated into the main project.

Before Git-style branching was introduced in commercial CAD software, researchers have considered branching's suitability for CAD models through examining the commonalities and differences between hardware and software development processes~\cite{Estublier2007}. Bricogne et al. and Fresemann et al. compared version control and change management in PDM (product data management) for the CAD design process to SCM (software configuration management) in the software development process. 
\color{black}
The major differences between hardware and software data management identified were: (1) software product versions are modified at the same time without affecting one another, but mechanical product versions are developed sequentially~\cite{Bricogne2012}; (2) mechanical product data is seldom modified after the product is released, whereas a released software product will likely change~\cite{Bricogne2012}; and (3) several developers modify a code version in parallel with the branch/merge approach, whereas only one CAD designer can develop a hardware product version with the check-in/check-out approach used in PDM (centralized version control)~\cite{Fresemann2021}. 
Bricogne et al. suggest that these differences between hardware and software data management contribute to the limited implementation of branch and merge in CAD~\cite{Bricogne2012}, and Fresemann et al. consider these characteristics in their design of a comparable branch-based PDM prototype inspired by SCM~\cite{Fresemann2021}.
\color{black}



Together these studies highlight the motivation for Git-style branching and merging mechanisms to support hardware development with CAD. Existing work has discussed potential usages and implementations of CAD branching but has not investigated practical use of commercial CAD branching tools by designers. Therefore, our study aims to determine the real user requirements of branching with commercial CAD platforms, the limitations of current branching functionality, and how to improve CAD branching tools to fulfill these user requirements. 


\section{Methods: Mining Online CAD Forums}


The goal of this study is to establish CAD branching use cases, and to identify the deficiencies of current CAD branching tools. 
We do this by mining user-generated posts from online CAD forums that discuss branching and merging and related activities for CAD, to build an understanding of what kinds of tasks CAD designers use branching for and what the limitations are to such tools. 
Through mining online forums, we analyze data from a large set of CAD users, and thus we can access a greater variety of discussions of branching use cases and shortcomings. 

Online CAD forums can be considered as a type of community of practice -- a ``group of people informally bound together by shared expertise and passion for a joint enterprise'' that facilitates learning, problem-solving, information exchange, and the sharing of best practices~\cite{Wenger2000} -- in this case, offering a place for CAD users to help each other solve design and development-related problems, share CAD knowledge, and discuss various CAD topics (e.g., drafting, modelling, simulations). Previous work by Kiani et al. has shown that online forums are the most frequently used resource among design professionals who use feature-rich CAD software, and the third most frequently used resource for CAD newcomers~\cite{Kiani2019,Kiani2020}. A survey study by Robertson et al. similarly found that ``experienced, constant users are more likely to visit online CAD forums'', stating that these users ``were keen to pass on the lessons they had learned from their years of experience''~\cite{Robertson2009}. Matejka et al. likewise found that CAD professionals are frequent online forum-goers -- nearly 25\% of their study participants visit forums daily~\cite{Matejka2011}. Evidently, online CAD forums are important sites for knowledge sharing, knowledge producing, and learning for novice and experts alike, and researchers can glean real user discussions, complaints, and requests that occur in such forums. 

\begin{figure}[ht]
  \centering
  \includegraphics[width=4.8in]{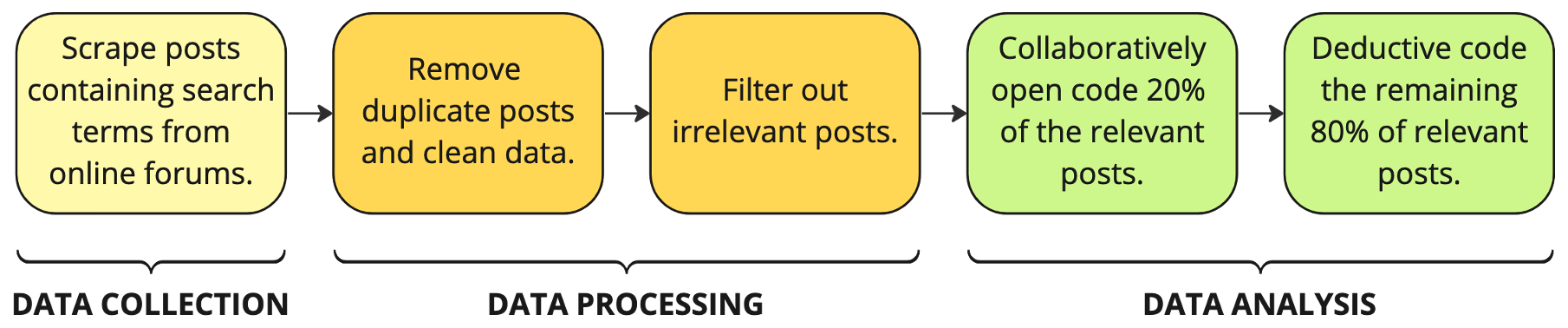}
  \caption{Overview of research methodology. Steps are coloured in yellow for Data Collection (Sec.~\ref{datacollection}), orange for Data Processing (Sec.~\ref{dataprocessing}), and green for Data Analysis (Sec.~\ref{dataanalysis}).}
  \label{methods}
\end{figure}

In this section, we first provide a brief overview of the studied forum sites. We then review the data collection and data processing steps, followed by the qualitative coding analysis used. Figure~\ref{methods} summarizes our research methodology. 

\color{black}
\color{black}

\subsection{Selection of Online Forums}
The five forum sites that we studied are described in Table \ref{forums}. It should be noted that these sites were not chosen at random; we purposefully selected online forums targeted towards CAD professionals or CAD enthusiasts who have a personal interest or motivation for discussing CAD-related topics. The web forums examined in this study include those officially affiliated with CAD programs (i.e., Onshape\footnote{https://www.onshape.com/en/}, SolidWorks PDM\footnote{https://www.solidworks.com/product/solidworks-pdm}, Autodesk Fusion360\footnote{https://forums.autodesk.com/}), as well as general, non-platform-specific forums (i.e., Eng-Tips\footnote{https://www.eng-tips.com/} and CAD Forum\footnote{https://cadforum.net/}). We chose Onshape, SolidWorks PDM, and Autodesk Fusion360 forums, specifically, because these CAD programs are the only ones that currently support or previously supported branching and merging~\cite{onshape2015,song2017,medeiros2017}. \color{black} 

When seeking online help, CAD users tend to visit product-owned forums over general forums because the terminology, construction workflows, and features can differ greatly between different CAD platforms, thus, product-owned forums offer more specific help~\cite{Kiani2019}. That said, non-platform-specific forums were also included in our study to capture general discussions of CAD version control that are not limited to a single platform.  

The platform-specific forums are moderated by employees of the company, whereas non-platform-specific forums are moderated by volunteers, who are typically professional CAD users or CAD enthusiasts.
All five sites allow public visitors to view posts, but in order to be able to further interact within the forums (e.g., create new posts, ``like'' comments, reply to others, and participate in a discussion), users must register as a member of that forum -- which involves creating an account that is verified with an email address -- to help ensure the quality and legitimacy of contributors. With our selection of forums, we aim to provide a representative sample of CAD designers who use or are familiar with branching. 

\begin{table}[ht]
\footnotesize
  \caption{Characteristics of the five sites studied, as of April 2022. Note that we are only able to report the number of contributors (i.e., registered users). The total number of forum users (i.e., registered, and non-registered users) may be higher.}
  \begin{center}
  \label{forums}
  \begin{tabular}{l p{2.5in} c c c}
      \hline
    Forum&Description (as stated on website)&Contributors&Posts&Creation Year\\
    \hline
    Autodesk&Documentation, tutorials, videos, and troubleshooting resources for Fusion360.&15K&120K&2013\\
    CAD Forum&A place for CAD users by CAD users. Discussion will center around CAD and all of its sources, resources, and results.&0.7K\color{black}&1.8K&2021\\
    Eng-Tips&Intelligent work forum for engineering professionals discussing engineering computer programs.&35K&300K&2000\\
    Onshape&Discussions about everything Onshape.&4K&18K&2014\\
    SolidWorks &SOLIDWORKS users discover, engage, and share about 3DEXPERIENCE Works -- SOLIDWORKS desktop, cloud-connected, and pure cloud.&3K&6K&2009\\
  \hline
\end{tabular}
\end{center}
\end{table}
\color{black}

\subsection{Data Collection}\label{datacollection}
In the online forums, participants discuss a wide range of topics, such as reporting bugs or requesting troubleshooting help from others. To narrow our search before data scraping, we searched for the following keyword terms (stemmed): 

\textit{branch*, merg*, copy*, clon*, fork*, git, github, version*}. We included the terms \textit{copy} and \textit{clone} because they are frequently used interchangeably with \textit{branching}~\cite{Linsbauer2021}. The term \textit{fork} was included because of its close relationship to branching in the software field, and the terms \textit{git}, \textit{github}, and \textit{version} to capture discussions of version control. 
\color{black}
For some forums, we could only search by one keyword at a time (e.g., a search for \textit{branch*}, then a search for \textit{copy*}), which resulted in duplicate posts. 

We built a custom web-scraper using Python (with packages \emph{Beautiful Soup}\footnote{https://www.crummy.com/software/BeautifulSoup/} and \emph{Selenium}\footnote{https://selenium-python.readthedocs.io/}) to extract the data from the forum posts which range from April 2005 to April 2022. 
Although commercial CAD platforms have only begun supporting branching since 2015 (by Onshape), we chose to include posts starting from April 2005 in our search criteria because this is when branching was introduced to the software development field by Git~\cite{Brown2018}; we found numerous posts discussing the potential of ``Git for CAD'', suggesting that CAD users have been asking for the ability to branch and merge long before its implementation in commercial CAD platforms. The data we collected include the title and content of the post, the initial posting date, the username of the post author, an ordered list of comments and corresponding usernames of commenters. Here we define \emph{post} as the initial posting and \emph{thread} as the entire discussion, which includes the initial post and any comments. The dataset was stored in a local Excel file for analysis.

\begin{figure}[ht]
  \centering
  \includegraphics[width=5in]{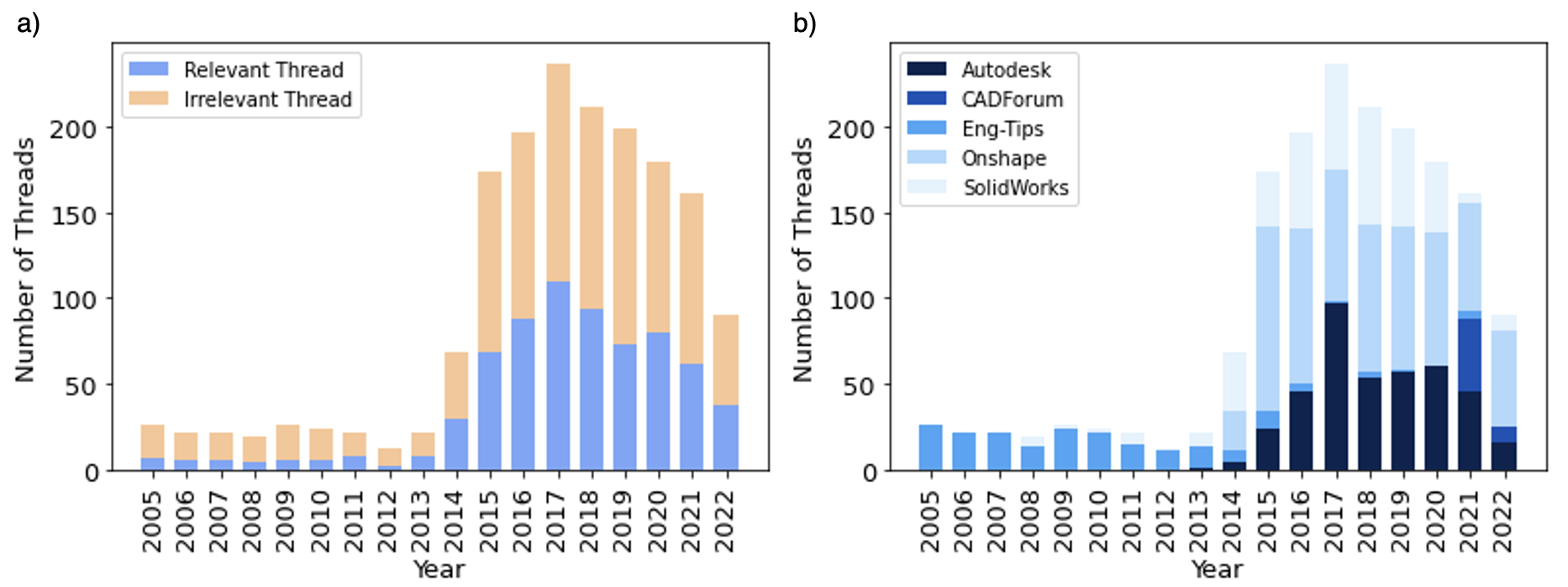}
  \caption{Characteristics of the dataset (n = 1,761): (a) Number of forum threads analyzed from April 2005 to April 2022, and (b) number of threads from each of the five forums per year. }
  \label{threads}
\end{figure}

In total, our search across the five forum sites returned 14,448 unique threads. Aggregate statistics for each forum are provided in Table \ref{forumstat}. The average length of a thread was 792 words (SD = 1,037). Over the 14,448 threads, there were 8,528 unique authors and the average number of posts per author was 1.7 (SD = 9.9). On average, each post had 8.6 comments (SD = 11.7) and 1,247 posts had no comments. The number of posts grew rapidly from 2005 to 2017 and the forum activity remained relatively steady from 2015 to 2022; the number of posts peak in 2017, which may be the result of Autodesk's launch and subsequent retraction of branching and merging in 2017 \cite{song2017}. Note that data collection in this work was done in April 2022, thus only four months of data was available for that year. Figure \ref{threads} shows some characteristics of our analyzed dataset.

\subsection{Data Processing}\label{dataprocessing}

After data collection, the dataset was cleaned, and duplicate threads were removed. As a first step in preparing the data, a random sample of 1,000 threads were manually parsed to identify keywords present in threads that were irrelevant to our research questions. Such keywords are presented in Table \ref{keywordfilter} and represent discussions about CAD software performance issues and specifications of computer hardware. Unlike other online review mining studies that included references to bug fixes, runtime errors, or glitches in their analysis, our study is interested broadly in how current branching features are used in CAD design and how to improve usability. Thus, we omitted threads that described unexpected bugs and glitches as their main topic, as these problems will likely be fixed by the developers in subsequent versions of the CAD software. The exception to this filter criteria were threads containing the word ``branch'' -- these threads were deemed as relevant even if they contained keywords like ``error''.

\begin{table}[h]
\footnotesize
  \caption{Types and examples of keywords that were used to filter out irrelevant threads from the dataset.}
  \begin{center}
  \label{keywordfilter}
  \begin{tabular}{l p{3.1in}}
     \hline
    Type of Keyword&Examples\\
    \hline
    Performance issues &``glitch'', ``crash'', ``slow'', ``bug'', ``runtime error'', ``error'', ``uninstall'', ``reinstall'', ``restart computer'', ``restart program''\\
    Computer specifications&``video card'', ``graphics card'', ``quadro'', ``nvidia'', ``firepro''\\
   \hline
  \end{tabular}
\end{center}
\end{table}

We prioritized filtering out irrelevant posts from the list of forum posts, knowing that this would result in a larger but more inclusive dataset, rather than to first filter by relevant posts and overlook potentially pertinent posts that did not fit our keyword search criteria. In this sense, if relevant posts are considered positive, we aimed to minimize false negatives and favour false positives. This was done because in the data analysis stage, manual coding and tagging was used, where the researchers could then determine the relevancy of posts with the help of additional semantic context. To give a confidence level of 95\% and confidence interval 5\%, we tested our keyword filter on a random sample of 400 threads, out of the total 14,448. Of the 400 threads, we found 0\% false negatives, 13\% positives (true and false, to be determined in the manual coding stage) and 87\% true negatives. After sample testing our keyword filter, the remaining threads were filtered for relevance to the research questions which returned 1,761 threads (around 12.2\%). 

\subsection{Data Analysis}\label{dataanalysis}
Once all 1,761 potentially relevant threads were extracted, the new dataset was manually coded by two researchers who have qualitative research experience and familiarity with CAD and CAD version control~\cite{Saldana2009}. 

A hybrid coding approach was used, whereby we started with the following broad categories for coding the data: (1) \emph{use cases:} how branches are currently being used; and (2) \emph{status of feature:} whether the branching use case is supported by existing functionalities. Within the \emph{use cases} category, we started with the high-level groupings of branching use cases found in software development literature (as presented in Sec.~\ref{sec_branchingsoftware}) to guide our initial three subcategories of product line management, risk isolation, and designer support -- though more subcategories could be added if they were deemed necessary. \color{black}
Codes within these three subcategories were then developed using an inductive approach, meaning that codes were added as new branching use cases were identified from the threads. A forum thread could be assigned more than one code if there was discussion of multiple use cases. 
Threads that did not discuss a branching use case were tagged by the researchers as \emph{irrelevant} and were discarded. 

The first two authors used the 80/20 approach to code the data; first, 20\% of the 1,761 threads (353 threads) were open coded collaboratively, then the remaining 80\% of threads were divided equally between the two researchers. 
\color{black}
According to the Institute of Educational Sciences (IES) guidelines, a minimum acceptable percentage agreement of 80\% must be achieved from 20\% of the total dataset \cite{IES2017} for the coding scheme to be deemed repeatable and consistent across all coders involved. Out of the 353 threads that were coded by both researchers, 83\% displayed complete agreement, 6\% displayed partial agreement, and 11\% displayed no agreement on any codes. Overall, the intercoder reliability percentage agreement was 88\% \cite{IES2017}. For threads with no agreement on any codes, one researcher deemed the thread irrelevant to branching whereas the other researcher considered the thread relevant and assigned the thread a branching use case. In this situation, the researchers revisited the thread and discussed until a consensus was reached. Of the 1,761 threads that were coded, 719 (40.8\%) relevant threads were found. Aggregate statistics for each forum are provided in Table \ref{forumstat}.

\begin{table}[ht]
\footnotesize
  \caption{Number of unique posts collected from the online forums (i.e., duplicates removed), number of posts after filtering by keywords, and number of posts relevant to the research questions after manual coding.}
  \begin{center}
  \label{forumstat}
  \begin{tabular}{l c c c}
      \hline
    Forum&Unique&Filtered&Relevant (After Coding)\\
    \hline
    Autodesk&2,839&405&150\\
    CAD Forum&398&51&17\\
    Eng-Tips&3,991&246&72\\
    Onshape&2,823&665&361\\
    SolidWorks&4,406&394&119\\
    \hline
    \textbf{TOTAL}&14,448&1,761&719\\
  \hline
\end{tabular}
\end{center}
\end{table}

\section{Results}

In this section, we first provide a detailed overview of our taxonomy of CAD branching activities, followed by an analysis of the shortcomings of branching tools. Where excerpts from the forum threads are quoted, we include the name of the forum site and the year it was posted in brackets; for example, a quote from an Onshape thread posted in 2021 would be followed by ``(OS21)'' -- ``OS'' for Onshape and ``21'' for 2021\footnote{Abbreviations for the five forum sites are: Autodesk (AD), CAD Forum (CF), Eng-Tips (ET), Onshape (OS), and SolidWorks (SW).}.

\subsection{Taxonomy of CAD Branching Use Cases (RQ1)}


To answer \emph{RQ1: Why is branching used for CAD?}, user-generated posts from online CAD forums were manually coded to identify and categorize branching use cases. Mimicking the three groupings of software branching use cases (as described in Section \ref{sec_branchingsoftware}), we found the similar three categories for CAD branching use cases: (1) \emph{Product Line Management} encompassing branching to manage the files and version of the mainline design, (2) \emph{Risk Isolation} used to experiment with changes without affecting the stability of the mainline design, and (3) \emph{Designer Support} to serve non-technical or organizational needs. The name of \textit{Designer Support} was adapted from \textit{Developer Support} to describe CAD users more accurately.

\begin{figure}[ht]
  \centering
  \includegraphics[width=5.5in]{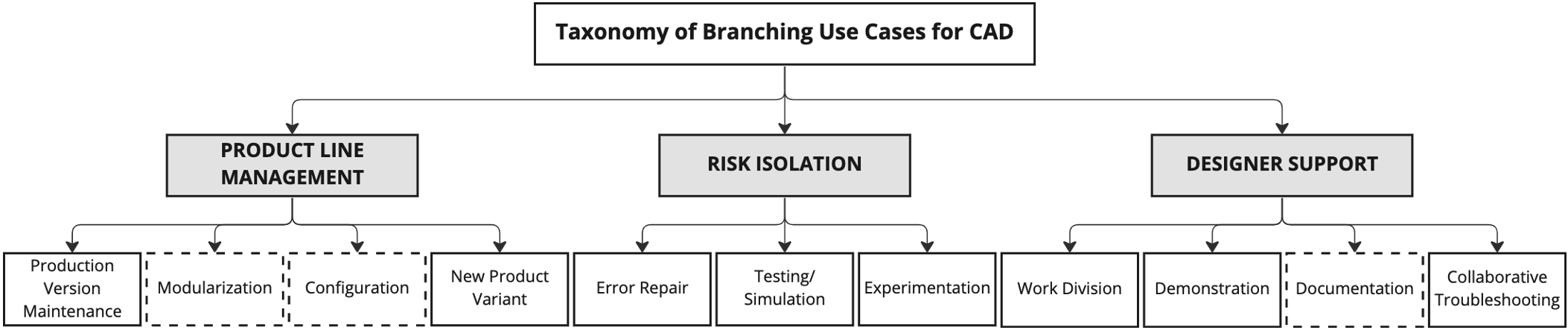}
  \caption{Taxonomy of CAD branching use cases. A dashed black border indicates that the use case is not an intended function of branching tools, instead branching is used as a workaround for another desired task. Consistent with our definition of use case in Sec.~\ref{sec_branchingsoftware}, each of the 11 use cases represent a distinct branching purpose that does not overlap with another, though a single branch could fulfill multiple use cases simultaneously.}
  \label{fig_example4.ps}
\end{figure}

\begin{figure}[ht]
  \centering
  \includegraphics[width=4.2in]{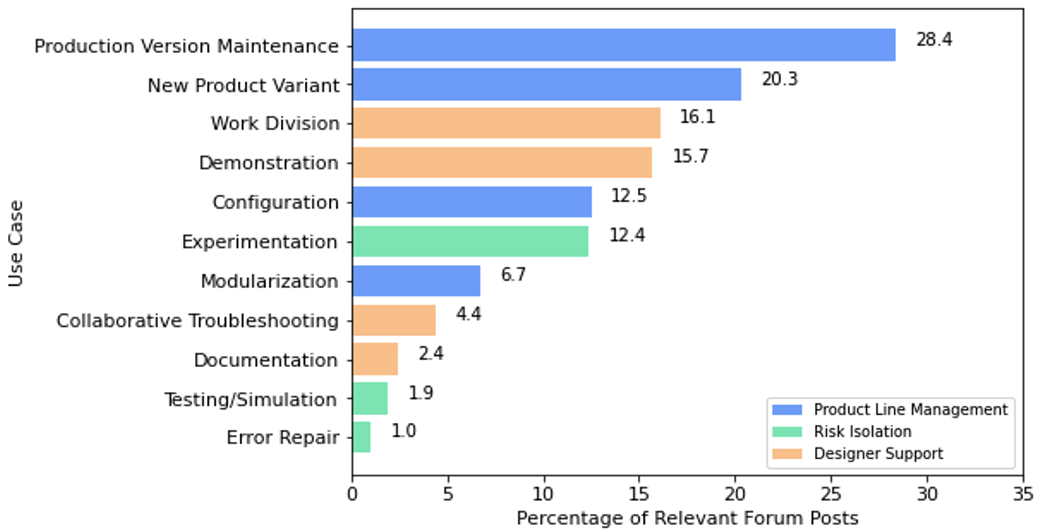}
  \caption{Frequency of mentioned CAD branching use cases in the forum threads (n = 719). \emph{Product Line Management} use cases are represented by blue bars, \emph{Risk Isolation} by green bars, and \emph{Designer Support} by orange bars. Note the percentages do not add up to 100\% because some forum threads discussed multiple use cases. }
  \label{codingfig}
\end{figure}

Figure~\ref{fig_example4.ps} shows the final taxonomy. 
\color{black}
To provide additional insight on designers' use of branching for CAD, we summarize the prevalence of use cases mentioned in the online forums, displayed in Figure~\ref{codingfig}. 


The identified CAD branching use cases are discussed in further detail below, and organized by the groupings of \emph{Product Line Management}, \emph{Risk Isolation}, and \emph{Designer Support}.

\subsubsection{Product Line Management}
\emph{Product Line Management} was identified as a top-level grouping of branching use cases in CAD, comprising all use cases that directly serve the mainline, release, or production version of the CAD file. In general, \emph{Product Line Management} branching facilitates the organization of a product's design process and evolution throughout the development pipeline, and thus, is highly related to the coordination aspect of collaboration~\cite{FUKS2005}. The branching use cases that were found for \emph{Product Line Management} are: production version maintenance, configuration, modularization, and new product variant. Software and hardware development share the common use case of \emph{Modularization} branching, but other use cases are unique to CAD. The \emph{Product Line Management} category is by far the most mentioned in the online forums (67.5\% of threads), and this is reflective of branching’s primary use in CAD as a version control tool.

\paragraph{\textbf{Production Version Maintenance}} 
\emph{Production Version Maintenance} branching encompasses any instance of branching used to separate the production version of the CAD design from any dependent files and to maintain the design’s currency and availability to designers. The production design version is the main branch that is intended to be revised, released, and then produced. As explained by a forum participant, \emph{``The main branch is where you put all the important changes to your model, and you only make changes to your main branch that are complete. The idea is that each version of the main branch is fully manufacturable, with no errors''} (OS16).

\emph{Production Version Maintenance} additionally encompasses using branching for version control, such as rolling back the design version to a previous iteration, branching from an older design, or storing design history. Regarding the use of branching for version control, one user explained, \emph{``I use merging/branching to save a version of my base part on the main branch, then create branches for additional features that I want to add to them. I try to make my branches precise and clean so that when they merge back in there are no issues''} (OS15).

When describing how branching is adopted in their workflow, another user wrote, \emph{``the easiest way to get some benefit from the branching system is to use it as a fancy undo, redo system. Whenever you are particularly happy with the state of a document-in-progress, save a new version. If you hit a dead-end in your design, open up the version control and branch off the last saved version that has the state you want to retry from. Rinse and repeat''} (OS16).

Of the 11 use cases, branching for \emph{Production Version Maintenance} was the most frequently discussed, mentioned in 28.4\% of forum threads.

\paragraph{\textbf{Configuration}} 
\emph{Configurations} are used to allow design parameters (e.g., component dimensions, quantity, and orientation of components) to be modified within a single design file to manage families of models, inclusive of parts and assemblies. Creating and managing configurations is commonly supported by CAD platforms and is achievable without the use of branching~\cite{Configurations2022}. Yet, \emph{Configuration} branching was frequently mentioned in the forum threads as a workaround that CAD users have developed in place of using the existing configuration functionality present in CAD platforms. Designers use \emph{Configuration} branches for many instances, such as to create two configuration branches of a design for two separate clients, or create a configuration to reflect each step in the manufacturing and machining process (e.g., investment casting configuration, then post-machined configuration). When choosing between these two methods of configuration control, there are trade-offs; as written in a forum thread, \emph{``The question is how many variations you have. If you only have a few, then branching/merging is a decent workflow. If you have 100’s of variations, then a table-driven approach seems more appropriate. In SolidWorks, this is called ‘table driven configurations'''} (OS16). Using table-driven automatically generated configurations is convenient when creating many permutations with few parameter changes. On the other hand, branching is effective for fewer, more exploratory design configurations, where many design parameters and features must be modified.

\emph{Configuration} branching was mentioned in 12.5\% of forum threads. Every major CAD platform supports configuration management without using branching; nevertheless, the frequency of threads discussing branching to serve the purpose of configurations suggests that for certain applications, designers consider branching to be more effective or efficient for creating configurations.

\paragraph{\textbf{Modularization}} \emph{Modularization} branching encompasses using branches to save different sub-systems or modules that can be combined through merging or assemblies to form a variety of different completed designs. \emph{Modularization} branching is frequently used for large and complex designs -- especially to coordinate interdisciplinary (typically, mechanical and electrical) designs that involve multiple interacting subassemblies with a high number of variable parts -- to allow for variability in module and component choice. One forum thread that described projects using \emph{Modularization} branching explained the scale that would warrant such practice: \emph{``My first couple of efforts were extremely complicated designs. One of them has some 50 components, another design has about two dozen components, but average about 20 to 30 bodies each''} (AD20). However, smaller, and simpler designs (comprised of few parts) rely less on \emph{Modularization}, because the simplicity of the design does not warrant dividing into modules -- \emph{Configuration} would be more suitable to allow for parameter variability.


In the relevant forum posts, branching for \emph{Modularization} (6.7\%) was mentioned nearly half as often as \emph{Configuration} (12.5\%), despite their similar objectives. However, considering that modularization is intended for models with high complexity and degree of variability between variations, the results suggest that simpler models are more common, or that branching tools are less frequently applied to highly complex and variable models.

\paragraph{\textbf{New Product Variant}} \emph{New Product Variant} encompasses using branching to build the basis for a new/separate design based on the original design. The high count (20.3\%) of posts relating to \emph{New Product Variant} is reflective of the popularity of the practice of reusing designs (which saves design time, allows for component interchangeability, and ensures a product's quality and reliability), and suggests that mechanical design best practices are actively employed by designers that use branching~\cite{Regli2011}. Creating a new product variant is analogous to ``hard forking'' in software development, whereby developers copy a repository with the intent to split off a new independent design that will not be later merged back to the original repository \cite{Zhou2020}. In CAD, designers use branching functionality to create new independent projects, even though hard forking and branching fulfill different development objectives. 
A further understanding of \emph{New Product Variant} branching in family-based hardware development may be informed by software literature regarding \emph{software product-line engineering}~\cite{Weiss1999}.

\subsubsection{Risk Isolation}
\emph{Risk Isolation} was adapted as a top-level grouping of branching use cases from software to hardware, encompassing all instances when branches are created to isolate unapproved or unverified file changes from the main branch of the design. \emph{Risk Isolation} additionally includes the use of branching to iterate new features and prepare CAD files for testing, as these are processes performed on files outside of the main branch. \emph{Risk Isolation} branching can be thought of as a cooperation tool because it enables designers to collaborate on shared tasks through a separate, yet linked workspace. Our taxonomy contains three use cases for \emph{Risk Isolation}: \emph{Error Repair}, \emph{Testing/Simulation}, and \emph{Experimentation}. Comparing \emph{Risk Isolation} branches in software and hardware, \emph{Error Repair} is analogous to \emph{Bug Fix}, \emph{Testing/Simulation} resembles \emph{Testing} branching, and \emph{Experimentation} serves a similar but nonequivalent purpose as \emph{Experimental/Prototype} branching. Of the three groupings, \emph{Risk Isolation} use cases are the least frequently discussed, present in 15.3\% of all relevant posts. 

\paragraph{\textbf{Error Repair}} Branching for \emph{Error Repair} in CAD is the process of creating a branch to fix a design error in a CAD file, with the intention of merging the file to the main design branch once the fix is complete, analogous to \emph{Bug Fix} branching in software design. 
Branching for error repair is performed to allow other design operations to take place on a CAD file while errors are being fixed, allowing the design workflow to continue uninterrupted during error repair. Error repair branching is typically performed at the project/assembly level, where the entire file is branched to fix an error. 
Regarding how to implement a branching approach to solve design errors, one forum user wrote: \emph{``if you make a change that breaks [the model], [...] find a spot on the history tree where it was working, version and branch from here and take a look at what's different''} (OS16). \color{black}

Despite being one of the most frequent uses of branching in software, branching to isolate error repairs from the main branch is uncommon in hardware design, mentioned in only 1.0\% of threads. This may be due to the difficulty or impossibility of selectively merging changes and error fixes to the main branch in CAD platforms, greatly reducing the utility of branching for error fixes. We further discuss this discrepancy between software and hardware branching for \emph{Error Repair} in Section \ref{sec_shortcoming_riskisolation}.


\paragraph{\textbf{Testing/Simulation}} Branches for \emph{Testing/Simulation} are created to run simulations, motion studies, and other CAE (computer-aided engineering) tests on a CAD model. Testing branches are used in two ways: (1) to run many different tests on the same design, or (2) to run the same test on multiple variations of a design and compare the results. Using branches to run different tests on the same model can be superior to other existing practices (e.g., creating many clones of the design or running the tests sequentially and resetting the model between tests) because branches can contain all of the tests within a single document and as put by one of the post commenters, \emph{``the branching method is much less `weight' than the full document copy method''} (OS19). One post author expressed a need to run an identical heat transfer simulation on an assembly with five different solid materials specified for one of the parts, while all other parts, parameters, and boundary conditions are held constant. By creating a branch for each of the five assembly variations, the simulations can be run simultaneously, and the designer can easily switch between branches to view and compare the results of each simulation. 

Branching for \emph{Testing/Simulation} was mentioned in 1.9\% of threads and is thus one of the least mentioned use cases.
 
\paragraph{\textbf{Experimentation}} \emph{Experimentation} branching involves the creation of a branch to serve as a ``sandbox'' for a designer to explore an alternative design, design and iterate new features, improve the design, or compare and select new features for inclusion in the production design. In the product design and development process, several candidate models are created and explored concurrently to allow for comparison and enable the selection of a single optimal design. Regardless of whether branching tools are available on the CAD platform, hardware designers invariably create and compare several candidate designs, but branching additionally allows for the light-weight creation of each candidate design in a separate branch, and for streamlined comparison and integration of the selected design to the main branch. 
One forum author illustrates: \emph{``being able to share variations of a design and have them available if you need to change directions [...] beats manual copies and renaming or playing with references. Not all users understand file management well enough to keep it straight, but branch and merge is more controlled and doesn't require that level of know how''} (SW22). \color{black}

Discussion of \emph{Experimentation} branching is present in 12.4\% of relevant forum posts.
\emph{Experimentation} is identified by CAD literature and CAD branching guides as one of the expected use cases of CAD branching; branching functionality in contemporary CAD platforms is explicitly designed to support branching for iteration and experimentation~\cite{onshape2016}.


\subsubsection{Designer Support}
\emph{Designer Support} use cases are intended to serve organizational or non-technical needs, and were found to predominately facilitate communication between designers and other stakeholders. Branching for \emph{Designer Support} was frequently discussed in the forums, mentioned in 38.6\% of relevant posts. Through mining online forums, we have found four branching use cases for \emph{Designer Support}: \emph{Work Division}, \emph{Demonstration}, \emph{Documentation}, and \emph{Collaborative Troubleshooting}. Software and hardware share \emph{Work Division} branching as a common use case, as well as \emph{File Sharing}, which is similar to \emph{Demonstration} in hardware design.

\paragraph{\textbf{Work Division}} Branches for \emph{Work Division} are used to divide modelling work between teams or individuals, so multiple designers can work on the same design concurrently. Branches provide each designer with their own individual workspace so that they can work without being disturbed by potentially distracting actions from collaborators, like hiding/showing parts, suppressing/un-suppressing features, or moving parts out of view. Using branches, collaborators can also access a CAD file without needing for the file to be ``checked-in'' to a central version control system. In addition to facilitating concurrent design, \emph{Work Division} branching is also used to coordinate a design's workflow -- for example, if the components of a product must be developed by multiple teams, branches are used to split and route design work between the teams.
An example of this is provided in a forum thread: \emph{``User 1 creates a branch and adds a hole. User 2 creates a branch and modifies the size of a cutout. Each branch gets merged back into the main workspace and updates the model at different times''} (CF21). \color{black}
In this regard, \emph{Work Division} branching is frequently used in combination with other branching uses, like \emph{Experimentation} and \emph{Modularization}, whereby each collaborator can branch to work on a different part or sub-assembly of the model or create their own alternative design. \emph{Work Division} branching was the third most mentioned use case (16.1\% of threads).

A specific use case within \emph{Work Division} branching is to support crowdsourced mechanical design. \emph{Crowdsource} branches are specifically used to allow a large group of, often geographically distributed, participants to contribute to the same design through the internet. Crowdsourced CAD projects typically rely on freelance designers, hobbyists, or general CAD enthusiasts to develop and maintain~\cite{Dortheimer2020,Eigner2017}. We elaborate on the use of branching in crowdsource mechanical design in Section \ref{crowdsource}.

\paragraph{\textbf{Demonstration}} \emph{Demonstration} branching is used to share and present a design to stakeholders. While related to \emph{Work Division} branching, the purpose of \emph{Demonstration} branching is to support design communication, whereas \emph{Work Division} branching is for coordinating design tasks and enabling parallel work. Throughout the development process, CAD models must frequently be accessed by many stakeholders, such as vendors, clients, external collaborators, manufacturers, or suppliers, for a variety of reasons, such as to provide feedback or approval, estimate a price quote, or receive design updates. Branching allows the designer to exact fine control over how the model is demonstrated; for a vendor, a designer may only want to share the parts of an assembly that they would like to be quoted and hide all other parts; for a client, protecting the intellectual property of a design is crucial, so a designer may create a branch to provide a ``clean'' version of the design that hides the design history (and parametric data necessary to recreate the model). In any case, branching also allows the stakeholder to view, manipulate, or annotate the design without disturbing the designer working on the main design branch. 
Using branching can also be preferred over other methods of sharing CAD designs (e.g., email) because it preserves \emph{``an audit trail of what model(s) were sent to the customer, at what Revision/Version''} (SW19).
\color{black}





\emph{Demonstration} is a frequently mentioned use case, appearing in 15.7\% of relevant forum posts even though it was not a branching use case that appeared in software literature. 
\color{black} So, the use of branching in this context implies that branching permits, at least for some designers, better or easier model demonstration to stakeholders than existing tools in CAD.

\paragraph{\textbf{Documentation}} Branching for \emph{Documentation} involves attaching and storing design documentation to a CAD file within a branch to improve the traceability of related files. From the forum threads, it was found that \emph{Documentation} branches are used to house a variety of files,
\emph{``from quotes to drawings to engineering documents, pictures, test results, user manuals, purchases, invoices, etc.''} (SW16).
\color{black}
\emph{Documentation} branching is a workaround (similar to \emph{Configuration} and \emph{New Product Variant} branching), as using branches to store design documentation is not an intended use case of branching and merging mechanisms, as far as we know. 

\emph{Documentation} branching was not a frequently mentioned use case, present in 2.4\% of threads. The low discussion of \emph{Documentation} could be attributed to the fact that branching tools were not designed to be used for this purpose~\cite{onshape2016}, thus designers are not aware of this workaround.

\paragraph{\textbf{Collaborative Troubleshooting}} \emph{Collaborative Troubleshooting} entails creating a branch with the intention of helping to improve or troubleshoot another person’s design. From the forums, we found that posts created to request assistance (e.g., deciding which type of mate to use, creating a new feature, or solving a design error) would often contain a shared link to the cloud-based model, in order to facilitate the communication of the design problem or error. With access to the post author’s design, commenters were able to view the complete model in its full context, as well as branch the design to propose direct and traceable modifications to the CAD model, without having to recreate the design problem on their personal workstation. The post's original author can then decide whether to merge or discard the changes. Naturally, finding this use case could be an implication of our methodology of mining and analyzing online forum posts, where some users go to seek help. However, branching to share knowledge has practical applications in industry hardware design teams. For example, a senior designer in a design firm can use branching to help troubleshoot a junior designer’s CAD model.

\emph{Collaborative Troubleshooting} branching was present in 4.4\% of forum threads. In all instances, the use case was mentioned in the comments of the thread, where a commenter branched the post author's design in response to their request for help. 
An example of such a response is: \emph{``Can you please either make your document public and post a link here, so we can take a look?''} (OS16).
\color{black}

\subsection{Shortcomings of Current CAD Branching Tools (RQ2)}\label{sec_shortcomings}

\begin{figure}[ht]
  \centering
  \includegraphics[width=4.5in]{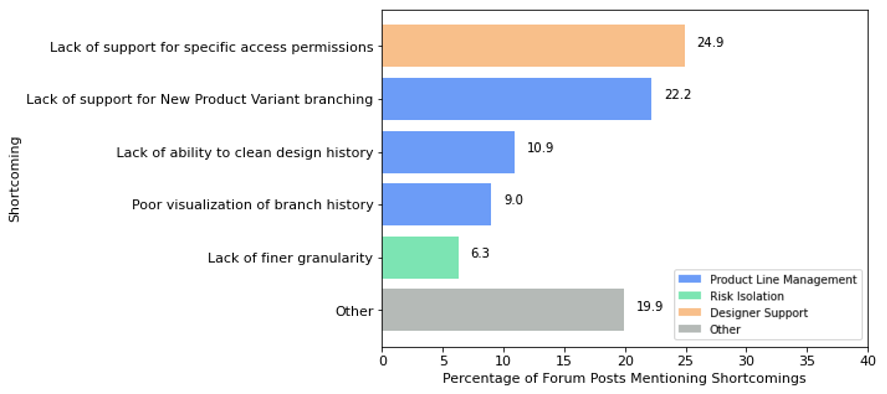}
  \caption{Frequency of CAD branching shortcomings mentioned in the forum threads (n = 221). Shortcomings related to \emph{Product Line Management} use cases are represented by blue bars, \emph{Risk Isolation} use cases by the green bar, and \emph{Designer Support} use cases by the orange bar. The grey bar represents \emph{Other} shortcomings which were less prevalent.}
  \label{codingfig2}
\end{figure}

The data mined from online forums provide important insights on why designers use branches, but also contain rich user discussions, in the form of criticisms and complaints, that are pertinent to improving future generations of CAD branching tools. In this section, we explore such user feedback and functional shortcomings, with the aim of answering RQ2 and proposing improvements to branching and merging functionality to better support collaborative hardware design. 
Of the total 719 relevant forum threads, we found 221 threads that mentioned a shortcoming of current branching functionalities. The prevalence of these shortcomings in the forums is shown in Figure \ref{codingfig2}. It should be noted that we only present the most commonly mentioned shortcomings; shortcomings that were uncommon or only described minor issues were tagged as \textit{``Other''}.

\color{black}

\subsubsection{Product Line Management}

Across the analyzed forum threads, three main shortcomings of CAD branching functionality for \emph{Product Line Management} were identified: (1) poor visualization of branch history, (2) lack of support for \emph{New Product Variant} branching, and (3) lack of ability to clean design history. 

\begin{figure}[h]
  \centering
  \includegraphics[width=5in]{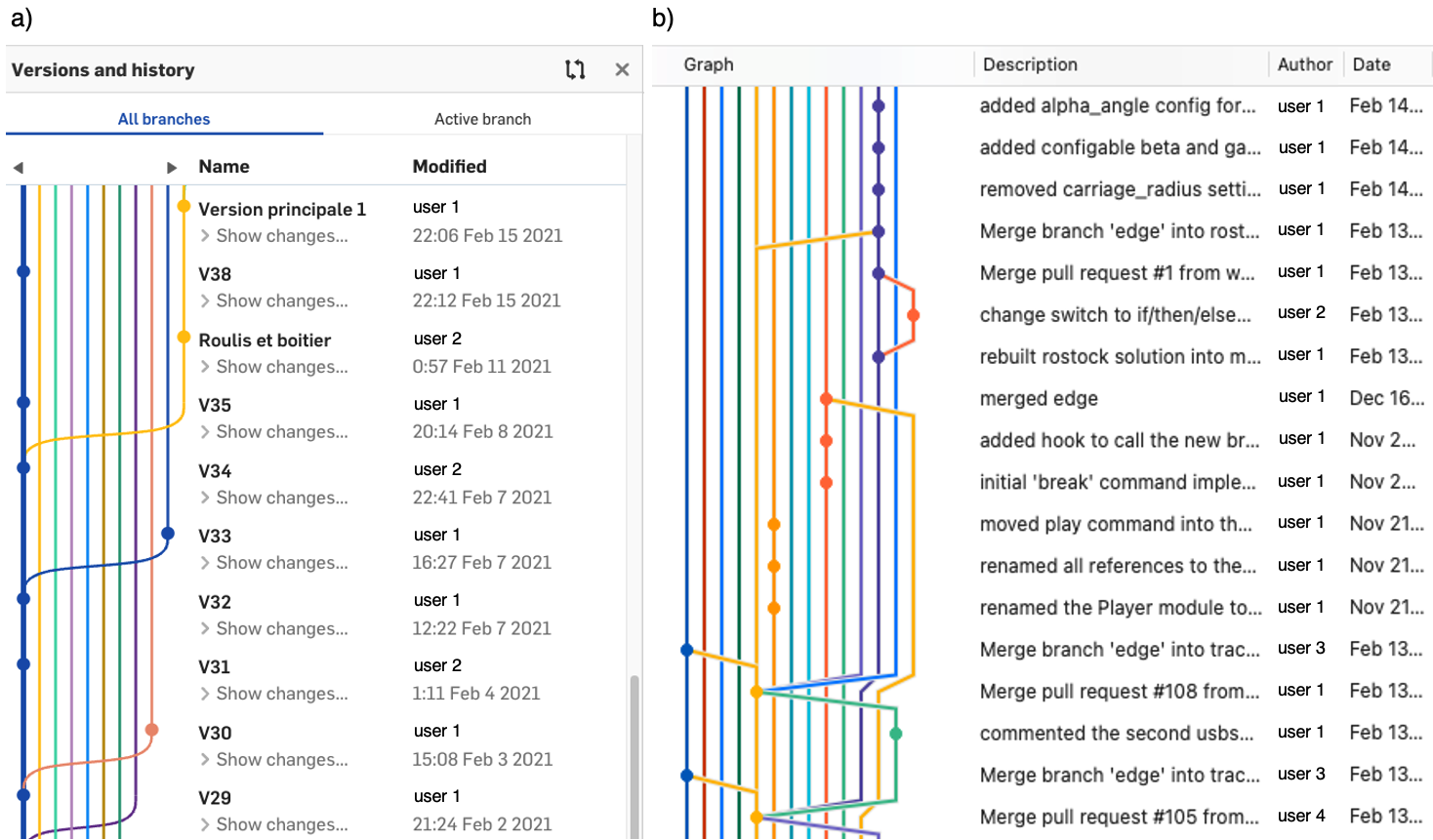}
  \caption{Comparison of CAD and software branch history visualization of: (a) Onshape's built-in graph, and (b) Git using Source Tree. For anonymity purposes, we have replaced the author names with ``user 1'', ``user 2'', etc.}
  \label{branchhistory}
\end{figure}

The most frequently mentioned shortcoming was that among those limited CAD platforms with existing branching functionality, there exists poor support of visualization and navigation of a model’s branching history, which can make it difficult for a designer to access and edit previous versions of a model. In software development branching tools (e.g., Git), the branching history of a codeline is available in the form of a commit history visualization, displaying branches and merges over time relative to a selected branch, shown in Figure~\ref{branchhistory}a~\cite{GitHub2022}. CAD users recognize the value of this functionality; several forum posts directly cite Git-style branching and commit history as a feature that would improve the utility of branching in CAD software. Although Onshape and briefly, Fusion360, have implemented a similar graphical user interface for designers to visualize a design’s branching history and activity (Figure \ref{branchhistory}b), navigating the graph is cumbersome and does not scale when the number of branches becomes large, and the commit history grows longer. The comment below illustrates:
\begin{quote}
    \emph{I’ve always had some sort of mental block with the Versions And History tree. The animation displayed for the SCROLLING VERSION GRAPH above is like a personal nightmare, complete with a sensation of vertigo while viewing. Usually I just avoid the tree and pretend it doesn’t exist, surviving on aptly-named tabs and documents. But that scrolling tree animation can’t be ignored: It jumps off the screen almost. I hope to comprehend this tree and its long vine-like branches eventually} (OS16).
\end{quote}

As shown in Figure \ref{branchhistory}, this challenge of poor visualization and navigation of commit history is present in both CAD and software development, where it is difficult for contributors to maintain an overview of forking/branching activity, and this lack of awareness problem has not yet been solved in either field~\cite{Zhou2018}.

The second major improvement request related to \emph{Product Line Management} is increased support for \emph{New Product Variant} branching. CAD designers commonly branch out an existing design to build the basis of a new, separate product variant, with no intention of merging back the variant. To support the creation of new product variants, designers need the ability to specify design aspects they would like to be carried over or discarded, such as comments, notes, or design history (versions and branches). Presently, branching to create a \emph{New Product Variant} only preserves the entire geometry of the model, and purges comments, notes, and design history.

Finally, CAD users have expressed a desire for a feature to prune the design history (versions or branches) of a CAD file in order to maintain a ``clean'' design, similar to the \emph{git-rebase}\footnote{https://git-scm.com/docs/git-rebase} feature. Predominantly, designers wish to use this feature to suppress \emph{Risk Isolation} branches, since these branches are often created for ad hoc reasons (e.g., fixing a design error or experimenting with a design change). As written in a forum thread:
\begin{quote}
    \emph{The notion of cleaning up a timeline reminds me very much of preparing a patch series in Git, in software development. I will often commit away until I’m happy with my changes, then rewrite the changes in a clean branch, creating a more logical commit series. That helps reviewers read the commits in order, which are now more like a story. Once they have made suggestions, I can often rebase them into that logical commit series and maintain the order for future readers} (AD16).
\end{quote}




\subsubsection{Risk Isolation}\label{sec_shortcoming_riskisolation}

A frequent complaint of existing CAD branching tools is the lack of granularity in branching. Current tools only support branching of an entire project file, but CAD users expressed the need to branch just a subset of the project, like a single part file or subassembly, similar to the \emph{Git shallow clone}\footnote{https://github.blog/2020-12-21-get-up-to-speed-with-partial-clone-and-shallow-clone/} feature. The need for finer granularity is particularly relevant for \emph{Risk Isolation} tasks because designers often want to experiment with a change or error repair of a specific component of the larger design and would like to do so without branching the entire project unnecessarily. Although current tools lack granularity in branching, selective merging (or \emph{``cherry-picking''}\footnote{https://git-scm.com/docs/git-cherry-pick}) at the part level has been introduced to CAD in 2022~\cite{Simon2022}, whereby a designer can choose to include or exclude changes from the source branch for each part in an assembly. Still, many forum posts discussed the need for even finer branching and merging granularity (e.g., at the feature level), such as for \emph{Experimentation} branching, where several candidate designs are being developed in parallel and designers want to merge a specific set of features of one candidate design (e.g., design A) to another (e.g., design B), while still maintaining other features of candidate design B.

\subsubsection{Designer Support}

In terms of the deficiencies of branching for \emph{Designer Support}, CAD users emphasized the need for more specific access permissions, which pertains to both \emph{Demonstration} and \emph{Work Division} branching. Since designs are shared for many reasons to many different stakeholders, it is necessary for designers to have tight control over the access permissions of shared designs. Forum participants discussed the ability to configure a variety of access levels, such as view only, copy only, edit, ownership, access to linked documents, and even automatic expiration dates for access privileges. Similarly, CAD users require the ability to assign branch- or version-specific permissions to stakeholders to protect intellectual property or prevent unwanted/unauthorized model edits. Currently, access permissions can only be set at the project level, and customized version control is needed to support CAD collaboration.


\section{Discussion}

\subsection{Hardware Versus Software Development Branching}\label{sec_hardwarevssoftware}
We begin our discussion with a comparison of hardware and software development, and how the respective characteristics of these related yet distinct fields impact the design and use of branching technology. 

\subsubsection{Alternative Designs}
Our study found evidence that CAD designers frequently discuss using branches to create and develop several alternative designs in parallel, with the intention of selecting one to put into production and discarding the rest, which we call \emph{Experimentation} branching. 
Typically in software development, the developed code is continually refactored, enhanced, and improved, but in the hardware development process, generating alternative designs to ultimately choose one is standard practice~\cite{Ulrich2015}. A possible explanation for this is that developing traditional software products is a fundamentally different process from developing hardware products. For physical products, 40\% of the development cost is attributed to physical prototyping and manufacturing \cite{Abdalla1994}. Due to the significance of these two phases, a great deal of development effort is spent ensuring that the optimal design is created before physical production, as it is much harder to change a physical product after it is released~\cite{Bricogne2012}. On the contrary, improving a software product after release is comparatively easier and faster, since a developer can still branch the codebase to deliver patches to the end user in order to fix bugs or add/remove a feature with the support of continuous integration, continuous delivery, and continuous deployment (known as CI/CD~\cite{Shahin2017}), thus it is less vital that the initial released design is the best design.

While it is common for software developers to iteratively improve a single design, other programming tasks (e.g., data science, building machine-learning models) involve experimentation with a variety of alternative designs, which is the practice of ``exploratory programming''~\cite{BethKery2017}. 
To support exploratory programming for data scientists, researchers have proposed different file structures to effectively express, compare, and visualize design alternatives, such as a side-by-side layout of alternative call paths within a single computational notebook~\cite{Weinman2021}. Although the new structure was helpful for design comparison, Weinman et al. also highlighted that more work is needed to make the interface more sophisticated and suitable for complex, real-world applications that may involve many more alternatives. This view is shared by Kery et al., who writes that there is currently a lack of tool support for experimentation, including a lack of support for recording and sensemaking of exploration history and a lack of support collaborative exploration~\cite{BethKery2017}. Another study by Kery et al. found that branching is seldom used for alternative design exploration in software contexts~\cite{Kery2017}, but as our findings have shown, experimentation branching is common among CAD designers. Evidently, tooling support for comparing alternative designs is an evolving challenge in both software and hardware development, and future studies should investigate how best to support design exploration.
\color{black}

\subsubsection{Branching for Risk Isolation}\label{sec_riskisolation}

An unanticipated finding of this study was that branching to fix design errors is not frequently performed or desired by CAD designers, with discussion of \emph{Error Repair} branching present in only 1.0\% of the relevant forum threads. This is contrary to studies of software development branching, where bug fixing is found to be one of the most common uses of branching \cite{Zou2019,Phillips2011}. There are several possible explanations for this result. For one, it is possible that errors are fixed directly in the production design version, as they are of sufficiently high priority to take over the entire design evolution -- in which case, this would apply to \emph{Production Version Maintenance} branching. Another potential cause is the multiple levels of extensive review that engineering firms require for CAD models before their inclusion into the main assembly file, meaning that design errors are identified and rectified in the initial component design and testing phases~\cite{Subrahmanian2015}. Finally, the infrequent usage of branching to fix design errors could be a consequence of CAD software’s 
lack of finer branching granularity (discussed in Section \ref{sec_shortcoming_riskisolation}) which may prevent designers from being able to seamlessly branch and fix an error within an individual component, then merge the fixed component back to the main assembly.

One of the new branching use cases found in our study is \emph{Testing/Simulation} branching, which has not been previously mentioned in CAD literature. Although \emph{Testing} is a known use of branching in software development \cite{Phillips2011,O'Sullivan2009}, testing hardware products is a more complex undertaking, as hardware products must be tested for multiple dimensions (e.g., thermal, fluid flow, materials, finite element analysis) that are not present in software design. A fundamental distinction is that in the software field, the developed code is the final released product. For hardware development, however, the CAD model is a representation of the product during development, and not the end product itself. Thus, running simulations and testing on a CAD model is an estimation of how the physical product will perform in real life, but the only way to definitively evaluate a product’s performance is to physically build and test it -- which is a significantly costly and time-consuming process~\cite{Li2016}. Given the importance of testing designs before fabrication and the utility of branching for this purpose, CAD users may find \emph{Testing/Simulation} branching to be a useful introduction of branching to their design practice.
\subsection{Branching as a Tool for Collaboration}\label{threeC}


In this work, we introduce a taxonomy that categorizes CAD branching use cases into the high-level groups of \textit{Product Line Management}, \textit{Risk Isolation}, and \textit{Designer Support}. Through studying these groupings and their use cases for branching, we are advancing a fundamental understanding of CAD collaboration, as defined as coordination, cooperation, and communication (i.e., the \textit{3C model} of collaboration~\cite{Ellis1991}). 

In CSCW literature, coordination is defined as ``the management of people, their activities and resources'' to facilitate cooperation and communication~\cite{FUKS2005}. The need to coordinate is extremely critical to the CAD design process; many individuals and teams share potentially 1000s of CAD files, and all prior file versions must be carefully tracked and managed~\cite{Cheng2023}. With the ability to branch, CAD users can improve coordination, via: organizing a design's evolution (\textit{Production Version Maintenance} branching); coordinating shared and individual tasks (\textit{Work Division} branching); splitting a large task into smaller tasks (\textit{Modularization}); and initiating a new goal to work towards (\textit{New Product Variant}). Given that branching has the potential for improving coordination in CAD collaboration, we envision that branching is an important feature that should be brought to the forefront for not only CAD designers, but for other stakeholders, such as project managers. Managers may find that a branch-based workflow can help organize personnel, timelines, resources, and tasks for product development projects.

Similarly, branches can enable cooperation (defined as ``the joint production of members of a group within a shared space''~\cite{FUKS2005}) in CAD work. With branching, designers can work in separate branches or ``copies'' of a CAD file concurrently, which allows tasks like \textit{Error Repair}, \textit{Testing/Simulation}, and \textit{Experimentation} to occur without impeding others who are also working on the same model in a parallel branch. 

Finally, our findings support that branching can help CAD users with the third aspect to collaboration, communication (defined as ``the exchange of messages and information among people''~\cite{FUKS2005}). For example, creating a branch for \textit{Demonstration} allows a CAD model to act as a shared artifact between individuals that can transfer a variety of different information about the design (e.g., geometry, material, size, construction sequence). CAD models are considered as virtual prototypes that have the ability to represent designs in varying levels of fidelity, and they are essential to effective communication throughout the product design stages~\cite{Camburn2017,Budinoff2022}. \textit{Collaborative Troubleshooting} branching helps to make a design error common among individuals, which allows them to achieve a shared understanding of the problem~\cite{Twidale2004}. To facilitate asynchronous communication in the CAD workflow, \textit{Documentation} branching is a way to preserve design-relevant information, such that it is accessible to other stakeholders in later stages of the development lifecycle.


Our results provide a framework and foundation for future work to advance theory and practice in branching of CAD artifacts. However, our work also has practical implications for the broader CSCW community. The mechanisms of branching and merging are not only used by software and hardware developers; we conjecture that the lessons learned in our study about the use cases for branching are highly applicable to other fields, particularly in graphical design, such as UX (user experience) design. Virtually any collaborative work field requires the management of prior versions of artifacts, co-development of intellectual goods, and the ``divide and conquer'' strategy enabled by branching and merging \cite{Sterman2022,Phillips2012}. Similar to our study, researchers in the CSCW community recognize the collaboration implications of branching and have made recent contributions regarding the application of branch-based version control to Scientific Workflow Management Systems (SWfMSs)~\cite{Abediniala2022}, open collaborative writing~\cite{PeThan2021}, and creative practice~\cite{Sterman2022}. With these and future insights, we hope to broaden the applicability and understanding of version control branching to improve collaboration efficiency across all CSCW domains.



\subsection{Supporting Crowdsourced and Open-Source Hardware Development}\label{crowdsource}

Our study found that CAD users are interested in using branching technology to support crowdsourced mechanical design. The term ``Crowdsourcing'' is defined as ``the act of a company or institution taking a function once performed by employees and outsourcing it to an undefined (and generally large) network of people in the form of an open call'' \cite{Howe2006}. Crowdsourcing has become a popular workflow in many design-centric CSCW domains, such as collaborative writing~\cite{Feldman2021}, web design~\cite{Oppenlaender2020}, graphic design~\cite{Foong2017}, and software development~\cite{Retelny2017,deSouza2020}. In software development, crowdsource branching is used as a way to easily delegate work amongst distributed collaborators which can reduce a product’s time to market, reduce the cost of development, leverage experts in the field, democratize and lower the barriers to participation, and offer invaluable learning opportunities for contributors~\cite{LaToza2016,li2015}.

The world of crowdsourced hardware development is much less mature than crowdsourced software development~\cite{Forbes2017}; yet our findings confirm that there is a growing interest in this area. In the forums, designers discussed recreational crowd design projects, such as a public CAD model of a virtual chess game that could be played when players create branches to move the chess pieces. Other, more formal products designed through crowdsourcing are intended to be put to production to solve a legitimate need. An example is Enabling the Future, an online global community of volunteers that uses crowdsourcing with Fusion360 to design and develop low-cost 3D-printed prosthetic hands for underserved populations~\cite{Schull2015}. 

When the scope of crowdsourcing is broadened, this is called ``open-source''. Open-source is a decentralized and collaborative method of development that not only allows for open contribution from an undefined and large network of people, but also gives the entire open-source community the freedom to study, use, modify, and share the product, maximizing the project’s beneficiaries~\cite{Olson2012}. The field of software development has been revolutionized by open-source (also known as open-source software or \emph{OSS}) -- creating world-leading products like Linux and Mozilla FireFox -- and this was largely made possible by the introduction of branching and merging functionality to distributed version control systems (DVCS) \cite{Fielding2005,li2015}.

The concept of open-source has also been applied to hardware development (OSHW), whereby the developed product is a tangible artifact that is free for the public to: (1) \emph{study} (i.e., the right to access sufficient information to understand how the product works and was designed), (2) \emph{modify} (i.e., the right to edit the product documentation and to create variants of the product), (3) \emph{make} (i.e., the right to physically fabricate and manufacture the product), and (4) \emph{distribute} (i.e., the right to share or sell the physical product or associated documentation) \cite{OSHW2022,Bonvoisin2017}. 


OSHW is rising in popularity -- attracting the attention of researchers, educators, and practitioners alike~\cite{Castro2019,Stirling2020}. However, the general sentiment towards OSHW is that while promising, the practical applications are presently limited~\cite{Eigner2017,Burnap2015}. The organizational challenges for adopting open-source in a hardware context are well-known, particularly with regards to protecting the intellectual property of designs~\cite{Herrera2020}, and establishing legal and regulatory frameworks to support open use, modification, and distribution of designs (which is more complicated than OSS due to the complex nature of hardware artifacts that may need to be described through 2D or 3D schematics, CAD files, bill of materials, or assembly instructions, instead of solely source code text)~\cite{Stirling2022}. However, even beyond organizational challenges, the expansion of OSHW is further limited by the shortcomings of current branching and merging tools (e.g., the lack of branching overview, lack of specific support for hard forking, inability to cherry-pick merges, and poor selection of access permissions, as we presented in Section \ref{sec_shortcomings}) and the lack of public, online platforms with sophisticated DVCS to support open-source CAD to the same extent as OSS~\cite{Bonvoisin2018}. 

In recent years, platforms for OSHW have emerged, such as Thingiverse\footnote{https://www.thingiverse.com/} and GrabCAD\footnote{https://grabcad.com/library} which play a significant role in ``Maker culture.'' This culture is an extension of Do-It-Yourself culture: a community of people interested in designing and creating physical objects. Makers often use CAD software and 3D printing~\cite{Oehlberg2015}, as it allows them to access and use CAD software and CAD models for free and to modify and share designs with the community
\cite{Novak2020}. Yet, while these platforms are sometimes viewed as the open-source version of CAD, they lack advanced tracking of file versions and dependencies (e.g., assembly relationships, hard forks)~\cite{Oehlberg2015}, leading to poor traceability, confusion caused by multiple versions of the same design, and low reuse of models~\cite{Oehlberg2015,Dai2020,Hofmann2018}. In a previous study by Bonoisin et al. that mined online hardware repositories, it was found that within a distributed hardware design project, there was significantly less collaboration on CAD files compared to other design-related files (e.g., documentation, images), suggesting that existing OSHW features are not sufficient to support the nuanced version control needed for open CAD collaboration~\cite{Bonvoisin2018}.
\color{black}

Although explicit discussion of branching for OSHW was not found in the forums, we conjecture that the growing interest in crowdsourcing hardware development is a promising step towards the possibility of OSHW. However, to accelerate a widespread and large-scale OSHW movement, the development of better branching functionalities and open-source CAD infrastructure is necessary.




\subsection{Limitations}
Here, we discuss the limitations of our study, first with respect to external validity, followed by internal validity.

\paragraph{External Validity} The first limitation to our work is that we collected data from English-language forum sites, so our findings of branching use cases and design shortcomings may not be generalizable to the global CAD user community. This lack of representation could have an impact on our findings of branching shortcomings (Section \ref{sec_shortcomings}), especially those that pertain to user experience and user interface; research has shown that \emph{cultural compatibility} is an important factor that influences a user's design preferences and interaction with a software system \cite{Alsswey2020}. 

Furthermore, we recognize that there may be a degree of bias in results derived from CAD platform-specific forums (in our case, Autodesk, Onshape, and SolidWorks), given that these sites are moderated by company employees who have their own incentives and biases. To address this, we also included non-platform specific forums (CAD Forum and Eng-Tips) in our data collection, where user feedback is more likely to be impartial. 


Lastly, there may be other valid use cases of CAD branching that were not found by us in the forums, either due to a limitation of our keyword search criteria, or simply the absence of public discussion of these use cases. Thus, it is in the interest of future studies to investigate branching practices in real-life design projects, as has been done in software research \cite{Zou2019}. The findings from our work serve as a promising first step in identifying and categorizing CAD branching activities. We also recognize that in this work, we do not claim that the identified use cases are best practices in CAD version control. However, with our developed taxonomy, a natural progression of this work would be to identify and develop best practices for branching in CAD, as has been similarly done in collaborative software development research \cite{Li2022}.

\paragraph{Internal Validity} Qualitative studies that rely on manual coding are susceptible to the researchers' bias. To mitigate this, two researchers coded the data with an 80/20 approach to reduce the effects of bias on our results from a single researcher. Additionally, a coding scheme was developed with detailed descriptions and specific examples for each code and sub-code to maintain consistency between the two coders. 



\subsection{Future Work: Merging in CAD}\label{sec_merging}

This paper aimed to explore branching intentions and purposes in hardware development. With this foundation, future work should study the related operation of merging, to advance our overall understanding of distributed version control in CAD. 

As previously discussed, merging was not the focus of this study because of the limitations of current CAD merging features (Sec.~\ref{sec_shortcoming_riskisolation}) and the dependence of merging on branching. Moreover, after our analysis, we found that the forum threads only minimally mentioned merging, and did not provide a comprehensive view of CAD merging practices (such as how often merges happen, how many branches are merged versus discarded, what prompts a designer to merge or discard, etc.). Nonetheless, given the importance of both branching and merging operations to software configuration management (SCM) and the need for better merging tools in CAD software, we believe that merging is a necessary area for future work.





In SCM, the version control system easily highlights changes between two branches or versions of code, identifies potential merge conflicts, and then merges branches by updating the root branch with the new lines of code made in the source branch. This process for merging is possible because software artifacts consist of lines of code expressed as text -- but CAD artifacts are different in structure~\cite{Frazelle2021}. 

Artifacts in CAD represent a physical object in 3D space using either a 3D point cloud (e.g., STEP files) or an ordered series of parametric operations that define the geometry of the object (e.g., SLDPRT files). As such, merging in CAD involves replacing an entire 3D point cloud with the new version of the point cloud (because selective merging at the feature level is not currently supported by any CAD software), or updating operations in a parametric model to implement the changed geometric features in the branch \cite{Koch2011}. Merging a 3D point cloud file without the ability to selectively merge features is redundant, as the part file is effectively being overwritten by the new version. 

In the case of a parametric CAD file, changes made in a branch might include those that remove geometries referenced by other elements. Consequently, a merge requires rolling back all changes to the point of divergence of the branch to allow the merge to be incorporated without conflicts, otherwise elements in the root component would reference geometries that have been changed or no longer exist~\cite{Sun2010, Jones2021}. The latter situation can create broken or impossible geometries that can render the entire file un-editable, and necessitates rolling back to a file version before the merge. 

Compared to the relatively straightforward operation of merging changed lines of code back into the root in software development, merging in CAD is complex, error-prone, and poorly supported by existing tools. As Frazelle writes in the article, \textit{A New Era for Mechanical CAD}, ``today there is no way to push a CAD file to a git repo, have several people modify the file, and resolve merge conflicts. (Well, maybe it could be done, but it would be the opposite of fun.)''~\cite{Frazelle2021}. 
\color{black} 
Thus, without improved functionality of the merging operation, the utility and appeal of branching will be severely limited for hardware designers. To fully realize the capabilities of CAD branching, future development needs to target merging functionality in CAD.

\section{Conclusion}

In this work, we conducted an empirical study of practical use of branch-based version control for hardware development with CAD. Through mining data from online CAD forums -- a rich and expansive source of user feedback -- we gathered over 14,000 user-generated posts across five forum sites, representative of over 8,000 unique CAD users. In the end, we analyzed 719 of these posts to build an understanding of envisioned and enacted branching practices in hardware design with CAD.

Our work contributes the first taxonomy of CAD branching use cases, as well as identification of actionable areas for improvement of current branching technology to better support modern product design and development. The taxonomy encompasses three high-level groupings of CAD branching use cases, \emph{Product Line Management}, \emph{Risk Isolation}, and \emph{Designer Support}, that parallel the groupings of branching use cases in software development, the field from which branching and merging originates. Our analysis of forum posts revealed that CAD designers not only use branching to fulfill technical design requirements, but branching is additionally used to support collaborative workflows and non-technical needs, such as designer communication, cooperation, and coordination. Moreover, branching was found to serve as a workaround for design tasks like creating design configurations, modularizing complex designs, and storing design documentation. 

We further identified and uncovered deficiencies of existing branching tools related to poor design history navigation, branching granularity, and configuration of specific access permissions, that may contribute to the slow adoption of this new functionality. Addressing these and other branching shortcomings is a fruitful area for future work and development. Future work will continue the investigation of CAD branching and merging in the context of real-life design projects, in order to confirm the branching use cases identified in this paper as well as expand our understanding of branching application in hardware development. Our work demonstrates that there is a real user need for robust Git-style branching and merging mechanisms within the collaborative CAD field, and the availability of efficient and effective branching and merging functionality in CAD could broaden the collaborative scale of hardware product design.

    

\bibliographystyle{ACM-Reference-Format}
   \bibliography{sample-base}

\received{January 2023}
\received[revised]{April 2023}
\received[accepted]{July 2023}

\end{document}